\documentclass[conference]{IEEEtran}
\IEEEoverridecommandlockouts
\usepackage{cite}
\usepackage{amsmath,amssymb,amsfonts}
\usepackage{algorithmic}
\usepackage{graphicx}
\usepackage{textcomp}
\usepackage{xcolor}
\def\BibTeX{{\rm B\kern-.05em{\sc i\kern-.025em b}\kern-.08em
    T\kern-.1667em\lower.7ex\hbox{E}\kern-.125emX}}

\usepackage[ruled,vlined,linesnumbered]{algorithm2e}
\usepackage{subfig}
\usepackage{float}

\begin{document}

\title{Managing Bufferbloat in Cloud Storage Systems\\
}

\author{
\IEEEauthorblockN{1\textsuperscript{st} Seyed Esmaeil Mirvakili}
\IEEEauthorblockA{
\textit{University of California, Santa Cruz}\\
Santa Cruz CA, USA \\
smirvaki@ucsc.edu}
\and
\IEEEauthorblockN{2\textsuperscript{nd} Samuel Just}
\IEEEauthorblockA{
\textit{Red Hat Inc.}\\
Los Angeles CA, USA \\
sjust@redhat.com}
\and
\IEEEauthorblockN{3\textsuperscript{rd} Carlos Maltzahn}
\IEEEauthorblockA{
\textit{University of California, Santa Cruz}\\
Santa Cruz CA, USA \\
carlosm@ucsc.edu}
}


\maketitle

\begin{abstract}
Today, companies and data centers are moving towards cloud and serverless storage systems instead of traditional file systems. As a result of such a transition, allocating sufficient resources to users and parties to satisfy their service level demands has become crucial in cloud storage. In cloud storage, the schedulability of system components and requests is of great importance to achieving QoS goals. However, the bufferbloat phenomenon in storage backends impacts the schedulability of the system. In a storage server, bufferbloat happens when the server submits all requests immediately to the storage backend due to a large buffer in the backend. In recent decades, many studies have focused on the bufferbloat as a latency problem. Nevertheless, none of these works investigate the impact of bufferbloat on the schedulability of the system. In this paper, we demonstrate that the bufferbloat impacts scheduling and performance isolation and identify utilizing admission control in the storage backend as an easy-to-adopt solution to mitigate bufferbloat. Moreover, we show that traditional static admission controls are inadequate in the face of dynamic workloads in cloud environments. Finally, we propose SlowFast CoDel, an adaptive admission control, as a starting point for developing adaptive admission control mechanisms to mitigate bufferbloat in cloud storage.
\end{abstract}


\section{Introduction}
\label{intro}
Scalability and availability drive the move to distributed and serverless storage systems.  Traditional file systems cannot compete when it comes to high performance storage that can tolerate common storage faults, and can scale across thousands of machines. Examples of such modern storage systems include the Google File System (GFS) \cite{ghemawat2003google} and the Hadoop Distributed File system (HDFS) \cite{shvachko2010hadoop}.

Distributed storage systems have become increasingly ubiquitous in modern information technology infrastructures. Despite their numerous benefits, these systems are not without challenges. One of the primary difficulties associated with distributed storage systems is the need to ensure a high Quality of Service (QoS) and fulfill Service-Level Agreement (SLA) demands in the face of mounting complexity, a growing number of components, and an expanding array of services. This challenge has been the focus of much academic research and industry attention in recent decades.

In order to achieve a comprehensive understanding of the challenges associated with resource management and scheduling in distributed storage systems, it is imperative to comprehend the general architecture and design of such systems. One of the most commonly used architectural frameworks in distributed storage systems is the Staged Event-driven Architecture (SEDA) \cite{welsh2001seda}, which enables the development of high-performance concurrent distributed systems while providing better resource management and schedulability through the control of request streams. The SEDA model defines the system architecture as a network of stages that are interconnected through event queues, offering fine-grained control over client requests. Notably, several distributed storage systems, such as Ceph \cite{weil2006ceph}, Cassandra \cite{lakshman2010cassandra}, and Amazon's Dynamo \cite{decandia2007dynamo}, adhere to the principles of SEDA.

Distributed and scalable storage systems that follow the SEDA principles consist of multiple components communicating asynchronously through queues. In most storage systems, a storage backend exists at the lowest level, responsible for storing the data structure on the storage devices \cite{aghayev2019file}. In order to store the data efficiently, the storage backend buffers the requests in a queue and flushes them onto the device in batches. Storing data in batches enables the storage backend to amortize the fixed overheads associated with single-write transactions, such as positioning, allocation, and metadata creation.

Request scheduling is a pivotal factor in achieving Quality of Service (QoS) objectives in a scalable storage server. The reason is that request scheduling can facilitate the implementation of essential QoS features, including fair resource allocation, guaranteed throughput and latency, and performance isolation \cite{yang2018principled}. A request scheduling mechanism can effectively allocate resources to various classes of requests based on their priority level, thereby ensuring different levels of service for different types of requests. For instance, requests generated by background data scrubbers are typically assigned lower priority compared to requests originating from a client.

The separation of request scheduling from the storage backend is a common preference among storage servers. Macedo et al. conducted a survey and classification of Software-Defined Storage Systems (SDS) \cite{macedo2020survey}, which defines an abstract SDS architecture and fundamental design principles. In their publication, they advocate for decoupling storage mechanisms from control policies over data and propose an SDS architecture comprising two separate layers: the Control Plane and the Data Plane. The Control Plane layer is responsible for system-wide control building blocks, while the Data Plane layer is responsible for storage operation stages. According to the SDS architecture, the frontend component, which represents the Control Plane, is the most suitable location for request scheduling.

Moreover, the majority of storage systems have multiple backends with varying designs capable of working with various storage devices such as HHD, SSD, and NVMe. Hence, if the request scheduling were to be placed in the backend, multiple scheduling mechanisms would need to be implemented to cater to the distinctive characteristics of each storage backend.

However, deploying the request scheduling on the frontend gives rise to a new challenge. The frontend must hold a sufficient number of requests in its queue to facilitate efficient request scheduling. At the same time, the backend must receive an adequate number of requests to operate efficiently with satisfactory performance. This leads to the fundamental question: what is considered sufficient for both the frontend and the backend? Balancing the number of requests in the frontend and backend presents two problematic scenarios.

The first scenario is when there are too many requests on the frontend and a few on the backend. In this scenario, while the scheduling mechanism has access to a significant number of in-flight requests, it is able to achieve its intended objectives without problems like priority inversion. However, the backend operations may suffer from poor performance as it becomes starved for requests.

The second scenario is when there are a few requests on the frontend and too many on the backend. A large or infinite queue on the backend without any back-pressure mechanism results in the second problematic case. This case leads to optimal performance due to the backend being inundated with requests. However, the frontend scheduling mechanism needs to work on fulfilling its objectives on a limited number of requests. The situation where excessive data is buffered downstream is commonly known as Bufferbloat \cite{gettys2011bufferbloat}.

Bufferbloat is a phenomenon that occurs in network systems where excessive buffering of packets takes place, causing significant delays and reducing overall network performance. In simple terms, it is a situation where the buffer in the network system is too large for the amount of traffic being transmitted. In the case of a bufferbloat, the buffer is not emptied quickly enough, leading to congestion and delays in packet delivery. 
    
    
    

We demonstrate how the phenomenon of bufferbloat is commonly observed in the context of storage systems, and can lead to concerns related to schedulability and Quality of Service (QoS). To mitigate this issue, it is necessary for the storage server to implement an admission control mechanism on the backend. Such a mechanism balances the number of requests between the frontend and backend. By ensuring a more balanced distribution of requests, the storage system can effectively address issues related to bufferbloat and improve overall performance and reliability.

Bufferbloat is a frequent issue in network systems. To mitigate bufferbloat, network experts have proposed various solutions, such as Random Early Detection (RED)\cite{floyd1993random}, Controlled Delay (CoDel) \cite{nichols2012controlling}, and Proportional Integral controller Enhanced (PIE) \cite{pan2013pie}. Given that storage systems can be treated as network systems, bufferbloat solutions in the network can be leveraged to implement admission control.

In this paper, instead of designing a new scheduling algorithm, we focus on improving the schedulability of the system by using admission control to mitigate bufferbloat. One of the main benefits of our work is to investigate the bufferbloat management mechanisms that can be easily adopted in existing systems without change in the system architecture or scheduling algorithm.

In this paper, we highlight the limitations of the static threshold-based admission control mechanism employed by some storage systems, such as Ceph, in addressing bufferbloat in the presence of dynamic workloads. To overcome this limitation, we propose SlowFast CoDel, a new adaptive algorithm designed for storage systems with varying workloads and storage devices. We leverage the CoDel algorithm, a well-known bufferbloat solution in network systems, as a starting point and develop the SlowFast CoDel algorithm, which can address storage systems' bufferbloat issues while adapting to changing workloads. We implement this algorithm in BlueStore \cite{aghayev2019file}, the storage backend of Ceph, and evaluate its effectiveness under different workloads.

Our primary contributions can be summarized as follows:

\begin{enumerate}
    \item We demonstrate that bufferbloat can lead to scheduling issues like priority inversion. To the best of our knowledge, our work is the first study on the impact of bufferbloat on the schedulability of queuing systems.
    \item We show that traditional static admission controls are inadequate in cloud storage, where the workload changes too often.
    \item We propose an adaptive admission control named SlowFast CoDel, which embeds a modified CoDel algorithm into an outer loop that dynamically adapts the CoDel parameters as workload changes. We introduce SlowFast CoDel as a starting point for bufferbloat mitigation in cloud storage backends.
    \item We implement our proposed SlowFast CoDel in Ceph's default storage backend, BlueStore, and evaluate its performance in mitigating bufferbloat in different workload scenarios.
\end{enumerate}


The rest of this paper is structured as follows. Section \ref{related} discusses related works. In section \ref{motiv}, we make a case for adaptive admission control to mitigate bufferbloat in storage systems showing that static admission control is inadequate in the face of dynamic workloads. Section \ref{dynamic} describes our SlowFast CoDel. In section \ref{eval}, we evaluate the implementation of SlowFast CoDel in Ceph BlueStore. Finally, we conclude in Section \ref{conclusion}.
\section{Related Works}
\label{related}

Over the past few decades, a significant body of research has been conducted on bufferbloat mitigation and admission control mechanisms. These studies may be classified into three distinct categories:
The first category concerns bufferbloat mitigation and concentrates on network performance enhancement. The second group contains the research on mitigating bufferbloat in storage systems. The third category pertains to admission control mechanisms and focuses on enhancing Quality of Service (QoS) and overall system performance.

\subsection{Studies on Bufferbloat Mitigation in Networks}
Research conducted in this group aims to address bufferbloat and congestion as issues pertaining to network latency. 

Sally Floyd et al. \cite{floyd1993random} proposed the Random Early Detection (RED) algorithm to manage congestion in packet-switched computer networks. RED operates by discarding packets before the network becomes congested, thereby preventing network congestion and reducing the average queue length. Despite being an older study, the paper's contribution is significant as the RED algorithm is widely adopted in modern computer networks to manage congestion.

Proportional Integral controller Enhanced (PIE) \cite{pan2013pie} is another algorithm that has been widely adopted in computer networks to address the bufferbloat problem. PIE reduces bufferbloat, improves network latency, and decreases packet loss by adjusting the queue size and packet transmission rate in response to changes in network congestion.

K. Nichols and V. Jacobson developed the Controlled Delay (CoDel) \cite{nichols2012controlling} algorithm to address the issue of bufferbloat in networks. The CoDel algorithm uses a simple, probabilistic approach to control queue delay. One of the key advantages of the CoDel algorithm is that it can respond quickly to changes in network conditions, such as sudden bursts of traffic or congestion, allowing prevent bufferbloat and improve network performance, even in dynamic environments.

Moreover, H. Jiang et al. \cite{jiang2012understanding,jiang2012tackling} identify and analyze the bufferbloat in cellular networks, which was not well-understood prior to this work. They propose a simple and easy to adopt solution that dynamically adjust the receive window. Their evaluation shows that the solution can reduce round trip time (RTT) and improve the throughput.

All these studies introduce bufferbloat solutions for network systems which have assumptions about the system that do not apply to storage systems. However, multiple studies have employed network abstractions and solutions in storage systems and have treated them as network systems. 

For instance, Stefanovici et al. \cite{stefanovici2016sroute} have presented a novel approach to designing a scalable and flexible storage stack by borrowing routing techniques from computer networks. The authors have argued that modern storage systems bear many similarities to computer networks, and one can exploit these similarities to enhance the storage stack's scalability, performance, and flexibility. Additionally, Thereska et al. \cite{thereska2013ioflow} have proposed a Software-Defined Storage (SDS) architecture that leverages the principles of Software-Defined Networking (SDN) to manage and regulate storage I/O flows. They have posited that SDS can benefit from SDN's programmability and flexibility.

\subsection{Studies on Bufferbloat Mitigation in Storage Systems}

Based on our current knowledge, the research conducted by A. Ravindran and colleagues \cite{ravindran2018edge} is the only study that specifically addresses the issue of bufferbloat in storage system queues and suggests a solution beyond network contexts. The study explores a suitable storage architecture for distributed vision analytics at the Edge. In their work, to enhance the signal-to-noise ratio (SNR), the researchers reduced the transmission rate when channel interference in wireless networks is detected. The study demonstrates that this approach can result in bufferbloat. To tackle this issue, the researchers proposed a control mechanism called "key-frame Sim" that discards a specific number of closest matching video key-frames from the buffers if the bufferbloat for video stream transmission exceeds a set threshold.

However, it is essential to note that key-frame Sim only addresses bufferbloat in the image and video stream flows for machine vision applications. Therefore, our research aims to design a mechanism that can address bufferbloat in all types of storage devices and workloads without the need for discarding data.

\subsection{Studies on Admission Control Mechanisms}
In storage server systems, where components are linked through queues, admission control and back pressure are considered to be effective mechanisms for regulating the flow of data and requests. In this section, we explore related works in data centers, databases, and web servers.

Goyal et al. \cite{goyal2022backpressure} introduce a congestion control mechanism named Backpressure Flow Control (BFC) for data center networks. This novel mechanism provides per-hop per-flow flow control with a bounded state and constant-time switch operations. The authors argued that end-to-end feedback protocols have become impractical due to limitations and current trends. The study demonstrated that BFC significantly enhances short-flow tail latency and long-flow utilization in networks with burst traffic. 

This work focuses on backpressure at the network level, aiming to achieve per-hop per-flow flow control in network switches. Moreover, this work requires changes in the networking protocol (sending and receiving special signal requests). SlowFast CoDel does not require any changes in the messaging system.

Cherkasova et al. \cite{cherkasova2002session} propose a mechanism for managing peak loads on commercial websites through Session-Based Admission Control (SBAC). The authors introduce two adaptive techniques, hybrid and predictive, that aim to maintain website responsiveness and stability. The hybrid admission control algorithm is designed to periodically switch between "strictly responsive" and "slightly less responsive" states based on the observed percentage of aborted requests and refused connections. On the other hand, the predictive strategy predicts the number of sessions the server can handle in a time interval and rejects any excess new sessions. The paper focuses on the session-based workload for web servers and proposes solutions to improve website responsiveness for clients. SlowFast CoDel does not require loss of service or request rejection as a signal to control rate.


Chauhan et al. \cite{chauhan2021optimal} introduce a novel approach to admission control and load balancing in a distributed real-time database system using deep reinforcement learning (DRL) and a memetic algorithm. The proposed technique utilizes local agents for admission control based on deep learning and a global agent to distribute admitted requests among the database nodes using the memetic load balancing algorithm. The design performs the scheduling of requests through these two types of agents, aiming to enhance system throughput and reduce latency. They evaluate the proposed approach through a simulation study, which demonstrates its effectiveness in improving system performance and reducing latency compared to other existing approaches.The paper implements the local agent admission control to accept or reject client requests based on the deep learning model and global agent feedback. However, SlowFast CoDel does not require service denial or request rejection. 

\section{Motivation}
\label{motiv}

In this section, we discuss the challenges of bufferbloat in storage systems and highlight the inadequacies of traditional admission control mechanisms to alleviate this problem in the context of dynamic and bursty workloads of cloud storage.

In the context of networks, the bufferbloat is the problem of latency. As requests are buffered in the large queues downstream, the latency of requests starts to rise. However, in storage systems, bufferbloat can lead to schedulability issues such as priority inversion. To the best of our knowledge, no research has been conducted on the influence of bufferbloat on the schedulability of storage systems.

Cloud storage systems rely on scheduling algorithms at various system levels to ensure tenant performance isolation. In storage servers (nodes), scheduling algorithms are employed to rearrange requests from the incoming queue and dispatch those with higher priority to the subsequent queue in the backend. However, if the backend has a significant buffer, the scheduler may tend to schedule and dispatch requests immediately, exacerbating the bufferbloat phenomenon. However, during a bufferbloat, the backend can increase the efficiency by committing larger batches to the storage device. Furthermore, storage backends employ their own scheduling algorithms to optimize the rearrangement of read and write requests, thereby maximizing storage locality.

On the other hand, bufferbloat can result in priority inversion in frontend scheduling. If the scheduler dispatches requests immediately, higher-priority requests that arrive subsequently must wait for lower-priority requests that were dispatched earlier. The frontend scheduling algorithm can prevent priority inversion and function more effectively with a larger pool of requests. As a result, inadequate admission control in the backend can compromise performance isolation among priority classes.

Within Ceph's object storage daemons (OSDs), the mClock algorithm \cite{gulati2010mclock} and its distributed environment adaptation, dmClock, have been implemented to ensure performance isolation between client (production), recovery, and best-effort workloads. This algorithm classifies requests into three aforementioned priority classes, each with a minimum resource reservation, proportional weight, and limit. The algorithm aims to assign each class its minimum resource reservation and, if system capacity permits, distribute the remaining available capacity among priority classes based on their proportional weight.

At first glance, it may appear that relocating frontend scheduling to the backend and removing the queue could present a viable solution. Nevertheless, incorporating high-level scheduling into the backend is not without its own set of challenges and limitations, which include:

\begin{enumerate}
\item As previously discussed in the Introduction, the most prevalent and effective approach is to decouple low-level storage mechanisms from high-level control policies. In general, frontend scheduling is responsible for ensuring that high-level quality of service requirements, which are typically defined at the system and business levels. Conversely, the storage backend is responsible for storing and retrieving data chunks efficiently on the underlying devices.

\item In cloud storage systems, the frontend component receives client requests and disassembles them into multiple storage operations, which are subsequently transmitted to the backend. For example, in object storages, high-level object requests are translated by the frontend into multiple metadata and chunk storage (read and write) operations. Furthermore, storage backends commonly feature their own scheduling algorithm, which reorganizes all chunk operations to enhance locality within the storage device.

\item Due to the heterogeneous nature of cloud environments, cloud-based storage systems are typically engineered to function with diverse storage devices and mechanisms, necessitating the ability to accept numerous pluggable storage backends. Consequently, moving the request scheduling mechanism, which is independent of storage device and technology, to the backend would not represent an intelligent decision.

\item Although designing a bufferbloat-free cloud-based storage architecture from the start is an interesting objective, many existing cloud storage systems are impacted by this issue. Therefore, this work focused on devising a straightforward mechanism that can be integrated into storage systems with minimal alterations to their architecture and scheduling algorithms. 
\end{enumerate}

In previous studies, the traditional admission control mechanisms usually aim to achieve better load balancing and prevent overloading components. However, in storage backends, we do not have load balancing or overloading problems. Load balancing is typically done at higher levels of the cloud before storage nodes. As a result, our focus is to design an admission control to control and alleviate the bufferbloat with the least impact on performance.

Employing traditional admission control systems can help the issue. However, these systems usually assume the requests follow a specific latency distribution that can be estimated and predicted. However, in the storage backend, many parameters and considerations can affect the IO request latency, including but not limited to:

\begin{enumerate}
    \item Type of IO: read or write
    \item Type of access: sequential or random
    \item IO size: 4KB, 8KB, 16KB, and ...
    \item Outstanding IOs: number of IOs currently in the buffer ahead of a request
    \item Type of the underlying device: HDD, SSD, or NVMe
    \item IO scheduling at the lower levels of the backend device or the operating system
\end{enumerate}

As a result, implementing an admission control in the storage backend that can work with different and dynamic workloads can be very challenging. This is the reason that many storage systems use an admission control with a static threshold. In such systems, the storage backend must be tuned for the incoming workload before deployment. For instance, Ceph has implemented a static budget-based throttle mechanism in its default backend, BlueStore. However, before deployment, the operation team should tune the throttle parameters (bluestore\_throttle\_bytes and  bluestore\_throttle\_deferred\_bytes) for the underlying device and environment workload \cite{ceph2023mclock}. 

However, when it comes to cloud environments where workloads can be very bursty and dynamic, such admission control systems tend to underperform the desired standards. As a result, it is crucial to design and use an adaptive admission control that can react and adapt to workloads and environment changes quickly.






\section{Designing a Dynamic Admission Control}
\label{dynamic}

\subsection{CoDel algorithm for Storage Systems}
\label{codel_for_storage}
The Controlled Delay algorithm monitors the queuing delay of the requests in defined intervals and decides how to handle the incoming flow. This algorithm has two key parameters, namely target delay and interval. The CoDel compares the minimum queue delay observed in every interval with the target delay. If the minimum delay exceeds the target, the CoDel initiates packet loss to make the upstream component decrease the TCP window and limit the packet flow. Moreover, with every violation, CoDel shortens the interval to adapt to the latency changes faster.

\begin{figure}[H]
  \centering
  \includegraphics[width=\linewidth]{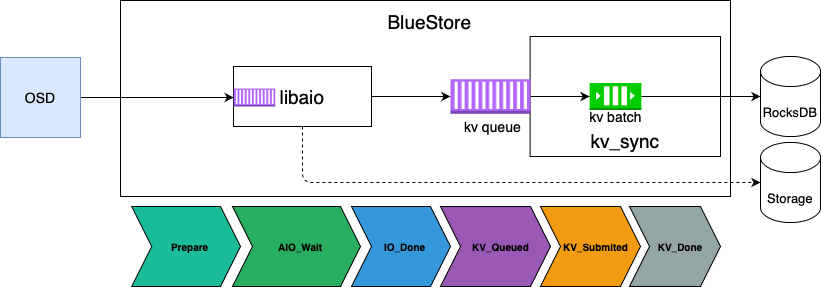}
  \caption{Requests path and states in the BlueStore from OSD (frontend) to the storage device.}
  \label{fig:bluestore_arch}
\end{figure}

However, we cannot use the CoDel algorithm as it is for the storage systems for the following reasons:

\begin{enumerate}
    \item The CoDel uses deliberate packet loss as a signal (to upstream) to control the packet flow. However, the components of a storage server are connected through queues. Consequently, we need a different flow control technique.
    \item The storage backend internals can be very complicated. For instance, as Figure \ref{fig:bluestore_arch} depicts, BlueStore consists of multiple components with their own queues. The IO chunks are queued to be stored by the asynchronous IO library provided by the OS. After that, the metadata associated with the IOs is queued in KV Queue to be stored in RocksDB \cite{dong2017optimizing}. As a result, tracking and measuring the queuing delay is not always feasible.
\end{enumerate}

For adopting CoDel in storage systems, we apply some changes in the CoDel algorithm. First, for request flow control, we utilize a throttle mechanism that limits the admitted requests to the backend according to a budget, which we call Queuing Budget. This throttle is similar to BlueStore throttle except that the queuing budget is controlled and adjusted by the CoDel algorithm dynamically. In other words, our modified CoDel algorithm controls the request flow to the backend by adjusting the queuing budget at every interval.

Another change to the CoDel is that we measure the backend's total latency instead of queuing delay to avoid complicating the CoDel algorithm and design a general solution for different backends. As mentioned before, we aim to design a solution that works with different storage backends and can be employed without too many changes to existing systems. We can achieve our goal by considering the backend as a black box by only measuring the total latency.

For the sake of clarity, from this point, we call this algorithm \textbf{Queuing Budget Adjusting CoDel}. Queuing Budget Adjusting CoDel records the total latency of the IOs admitted to the backend within an interval. At the end of the interval, if the minimum observed latency is higher than the target latency, it decreases the queuing budget in the throttle. Otherwise, it increases the queuing budget to admit more IO requests. Algorithm \ref{alg:codel} shows the process of Queuing Budget Adjusting CoDel.

\begin{algorithm}
\SetAlgoLined
backend\_queuing\_budget = initial queuing budget\;
 at the end of every interval\\
 \Indp
 \textbf{begin}\\
 \Indp
 \eIf{$min\_latency >$ \textbf{TARGET\_LATENCY}}{
 Decrease $backend\_queuing\_budget$ according to $|min\_latency - target|$\;
 }{
 Increase $backend\_queuing\_budget$ by $budget\_increment$\;
 }
 reset $min\_latency$\;
 sleep for \textbf{INTERVAL}\;
 \Indm
 \textbf{end}
 \caption{The Queuing Budget Adjusting CoDel for Storage Systems}
 \label{alg:codel}
\end{algorithm}

\quad

Queuing Budget Adjusting CoDel adjusts the throttle queuing budget based on the latency of the backend and a given target latency. However, the target latency parameter depends on the workload, meaning that different workloads need different target latency values to operate as desired. This means the algorithm needs parameter tuning for every workload to work at the desired performance state. However, when the workload is dynamic and unstable, having predefined fixed CoDel parameters is proved to be difficult.

In order to overcome this issue, we propose the SlowFast CoDel, an adaptive dual-loop control mechanism. In the core of SlowFast CoDel, Queuing Budget Adjusting CoDel adjusts the queuing budget in every interval (Fast Loop), and a slower loop with longer intervals, called \textbf{Target Adjusting Loop (Slow Loop)}, adjusts the target latency of the Fast Loop.

\subsection{SlowFast CoDel}
SlowFast CoDel needs to adapt and optimize itself based on two variables namely "backend latency and throughput". Optimizing a system based on two variables can be a difficult task. Typical methods for such optimizations are Gradient Descent and Newton's method. In such methods, the algorithm tries to minimize/maximize an objective function (of variables) in iterative steps. In every step, the algorithm updates the controlling variables. 

There are many examples of using gradient descent in a wide variety of systems. For example, Wang et al. \cite{wang2021asynchronous} investigate an asynchronous gradient descent algorithm for resource allocation in distributed systems, and 
Kalanat et al. proposes algorithms to find optimized actions in social networks \cite{kalanat2020action, kalanat2020extracting}. However, using iterative gradient descent in storage systems can complicate the system and increase the computation overhead. Moreover, gradient descent algorithms are unstable in low-level systems like the storage backend. The reason is that storage backend behavior can be very noisy due to different factors such as background tasks, type of storage device, faults in devices, etc. 

\begin{figure}[h]
  \centering
  \includegraphics[width=\linewidth]{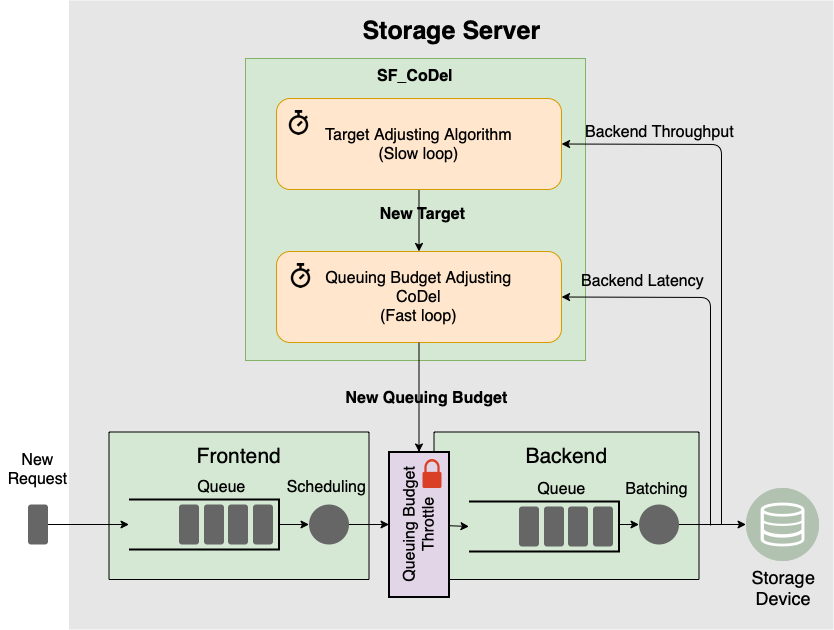}
  \caption{SlowFast CoDel algorithm consists of Queuing Budget Adjusting CoDel (Fast loop) and Target Adjusting Loop (Slow loop).}
  \label{fig:sf_codel}
\end{figure}

In such unstable noisy systems, a dual-loop control algorithm can work more efficiently with stability. In a dual-loop control algorithm, one loop tries to optimize one variable by adjusting the system's parameters, and a second loop tries to optimize the other variable by adjusting the parameters of the first loop. We designed SlowFast CoDel as a dual-loop control algorithm. SlowFast CoDel algorithm consists of two separate optimization loops, namely fast (inner) and slow (outer) loops. Figure \ref{fig:sf_codel} shows the structure of this algorithm.

The Fast Loop is the Queuing Budget Adjusting CoDel that monitors the backend latency and controls the backend queuing budget based on the target latency parameter. The fast loop quickly reacts to latency and workload changes by measuring and controlling the latency in high-frequency (short intervals). 

The Slow Loop (Target Adjusting Loop) is a low-frequency loop that monitors the backend throughput and balances the throughput loss and latency reduction by adjusting the target latency parameter in the fast loop over longer intervals. As the fast loop limit the admitted IOs to the backend, the backend latency and the throughput decrease. The slow loop changes and optimizes the target latency parameter with respect to the throughput-latency trade-off to distribute IO requests between the backend and frontend properly. 

As the throughput reaches the maximum capacity, the slow loop decreases the target latency to prevent bufferbloat. As the ratio of throughput loss to latency loss increases, it increases the target latency to prevent the backend from starving. Since the throughput of the backend is usually unstable due to the effect of external sources such as compaction or device behavior, a low-frequency sampling over a longer interval is the best approach to keep the optimization stable.

\subsubsection{Target Adjusting Loop}
\quad

Since the target latency is a workload-dependent parameter, Queuing Budget Adjusting CoDel algorithm is not able to adapt to different workloads with a fixed target latency. In order to manage varying workloads, it is necessary to adjust the target latency parameter in the fast loop. When the designated target latency is excessively low, the throughput can fall below the acceptable range. On the other hand, if the target latency is excessively high, the bufferbloat will occur.

\begin{figure}[h]
  \centering
  \includegraphics[width=\linewidth]{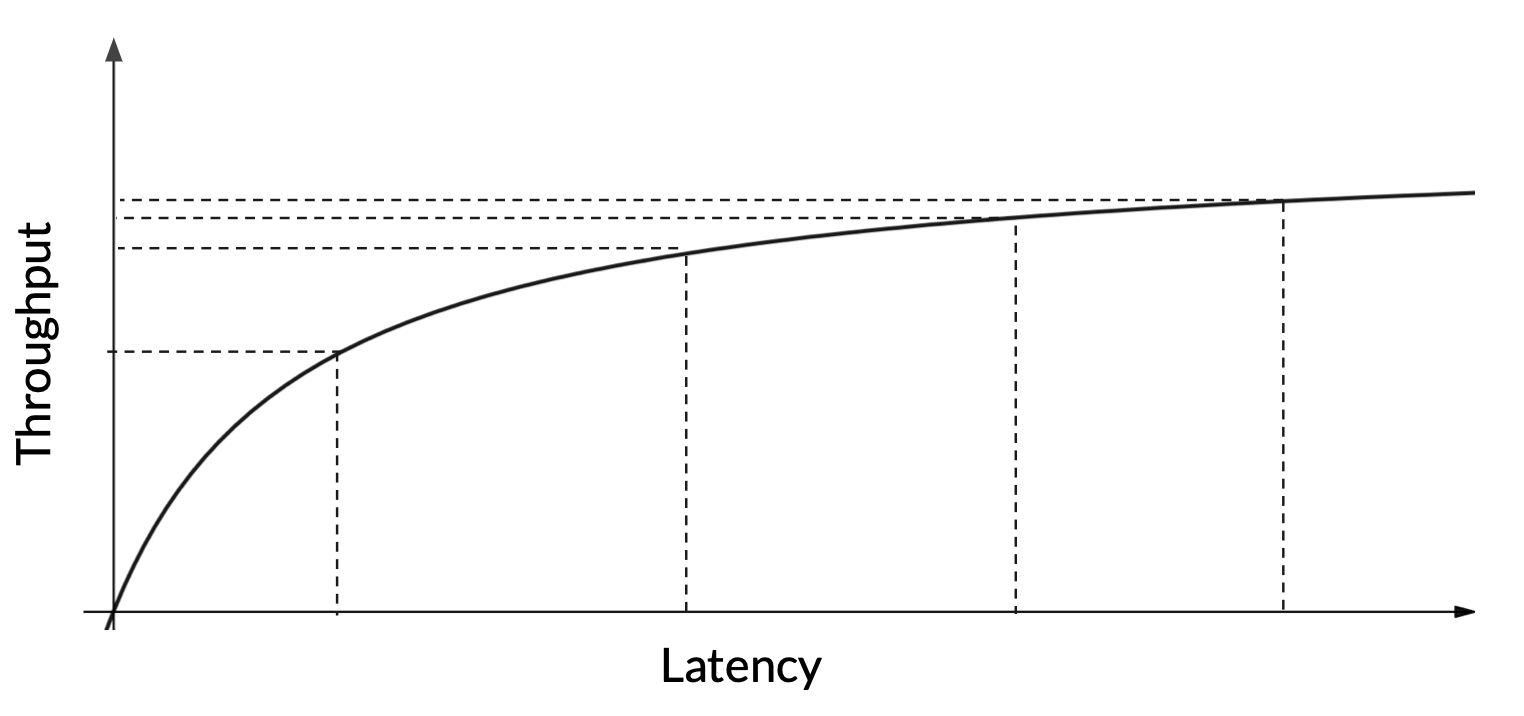}
  \caption{The throughput-latency curve in storage backend. By increasing the latency, throughput increases in logarithmic rate until it reaches the system saturation point.}
  \label{fig:throughpu_latency_curve}
\end{figure}

Target Adjusting Loop uses the Throughput-Latency curve to find the suitable target latency. Figure \ref{fig:throughpu_latency_curve} shows a typical Throughput-Latency curve for storage backends (and many other systems). This curve indicates that increasing the backend load, backend throughput, and latency will increase. However, at some point, increasing the backend load cannot improve the throughput beyond system capacity. However, the bufferbloat occurs from that point on since the backend is saturated.

The slow loop captures and monitors the throughput and latency history of the backend throughout a certain number of past intervals to estimate the Throughput-Latency curve. Using the estimated curve, the slow loop can select the suitable target latency to avoid bufferbloat and preserve an acceptable throughput. As the workload changes, the slow loop can choose a reasonable target latency according to the throughput-latency curve changes.

In the Throughput-Latency curve, the slope of the curve is a good indicator of bufferbloat regardless of workload size and type. If we pick a target latency where the slope of the curve is closer to zero, the impact of the bufferbloat will be higher. On the other hand, if the slope of the curve at the selected target latency is too high, the throughput will drop, and the backend will starve. To choose a proper target latency, the Target Adjusting Loop estimates the Throughput-Latency curve and picks the target latency at which the slope of the curve is equal to a fixed parameter called \textbf{Trade-off Parameter}.

As mentioned, target latency is a workload-dependent parameter. A workload-dependent parameter needs to be tuned and defined separately for different workloads. On the other hand, a workload-independent parameter performs its desirable effect on the system independent of what kind of workload the system operates on. An adaptive and auto-tuning system should be free of any workload-dependent parameters. Instead, it should provide some workload-independent parameters so that the user/operator can customize it based on demands and preferences. The Trade-off parameter in the Target Adjusting Loop is a workload-independent parameter that provides control over the throughput and latency trade-off. In section \ref{sens}, we explore how this parameter can affect the results for different workloads. This parameter does not need to be tuned for different systems.
\section{Evaluation}
\label{eval}
In this section, we show how bufferbloat can affect performance isolation in Ceph through experiments. After that, we show how admission control can mitigate bufferbloat effects in Ceph. Finally, we demonstrate the necessity of adaptive admission control to handle the dynamic workload in cloud environments.

\subsection{Experimental Setup}
We set up a Ceph cluster of four OSD nodes with an SSD storage device to perform our experiments. We used the latest release of Ceph to this date, Quincy, the 17th stable release. Moreover, we utilize the Ceph Benchmarking Tool (CBT)  to run our experiments' workloads. For the frontend (OSD) scheduling mechanism, we use mClock with the following policy.

\begin{itemize}
    \item Client: minimum reservation of 1000 requests and weight of two.
    \item Recovery: minimum reservation of 1 request and weight of one.
\end{itemize}

This policy should ensures that the recovery workload will not affect the client's workload.

In addition, we pick Trade-off parameter of 0.1 for our experiments. In section \ref{sens}, we explore the different options for this parameter.

\subsection{Bufferbloat Impact on Storage Schedulability}
\label{schedulability_impact}
In order to show how bufferbloat can affect storage schedulability, we devise the following test scenario. In our setup, first, we bring down an OSD and populate the other healthy OSDs with data while one OSD is down. Then, we bring the down OSD back up. When the recovery starts, we start a bursty client workload as the recovery is happening in the cluster.

In our setup, we set the mClock policy to limit the recovery to favor the client's workload. However, since the client workload is bursty, the OSD queue can occasionally be filled with only recovery requests. In such situations, mClcok dispatches recovery requests to the backend in the absence of requests with higher priority. However, when the client workload arrives, multiple recovery requests have been dispatched to the backend. Since our mClock policy is to prioritize the client workload, this is an priority inversion. With proper admission control, priority inversion in such cases can be manageable by preventing the frontend from dispatching too many requests.

\begin{figure}[h]
  \centering
  \includegraphics[width=\linewidth]{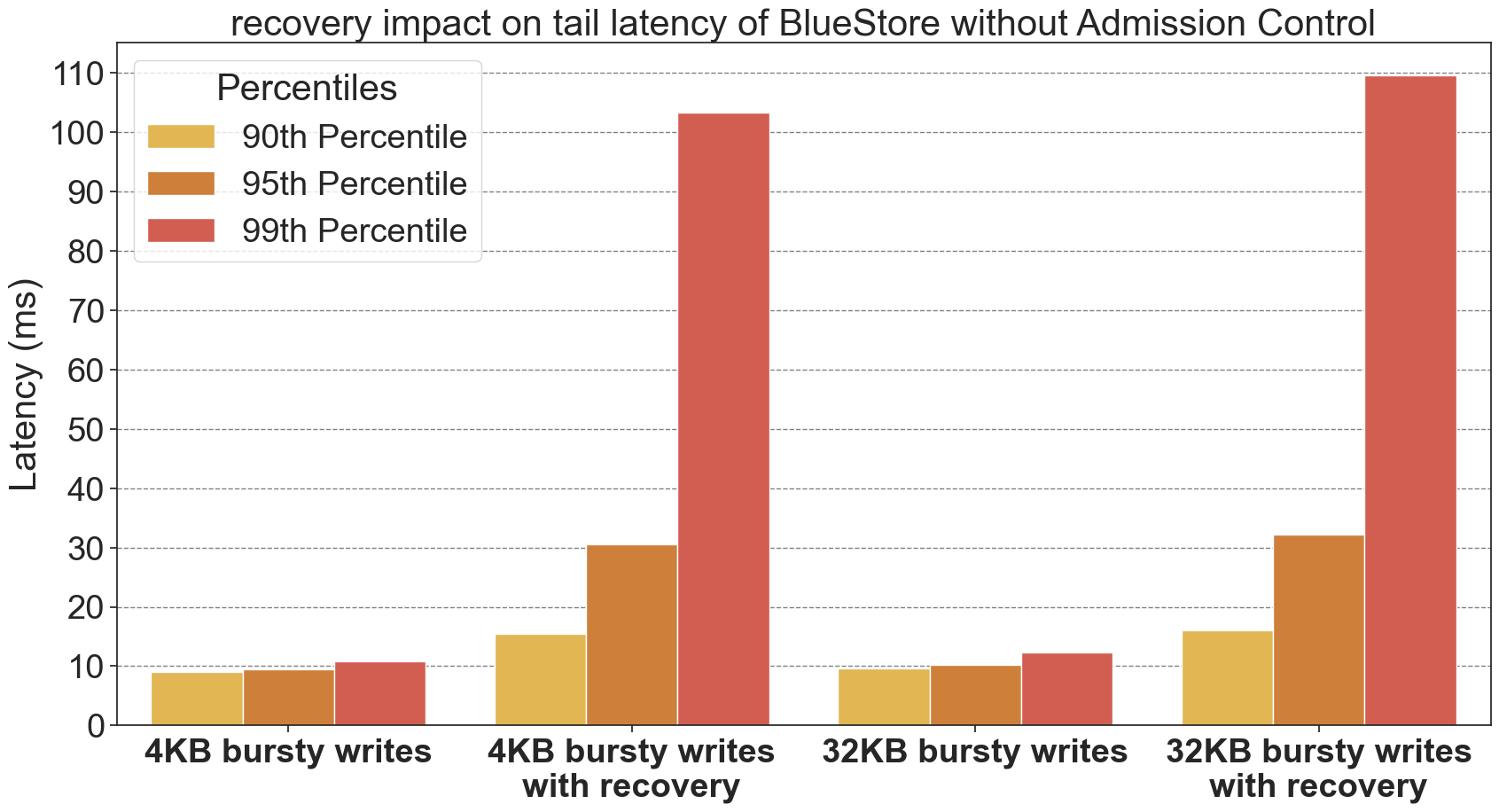}
  \caption{Impact of bufferbloat on client latency in absence of an admission control.}
  \label{fig:bufferbloat_impact}
\end{figure}

Figure \ref{fig:bufferbloat_impact} shows how recovery workload can affect the client tail latency in the system without any admission control on the backend. We run 4KB and 32KB bursty client workloads with and without recovery. As depicted, when the recovery is in progress, the tail latency of the client workload increases substantially especially 99th percentile. In the absence of a storage backend admission control, priority inversion can hurt a client's busrty workload.

\begin{figure}[h]
  \centering
  \subfloat[\textbf{4KB bursty writes} workload with recovery\label{fig:ac_bufferbloat_4k_64}]{\includegraphics[width=0.5\textwidth]{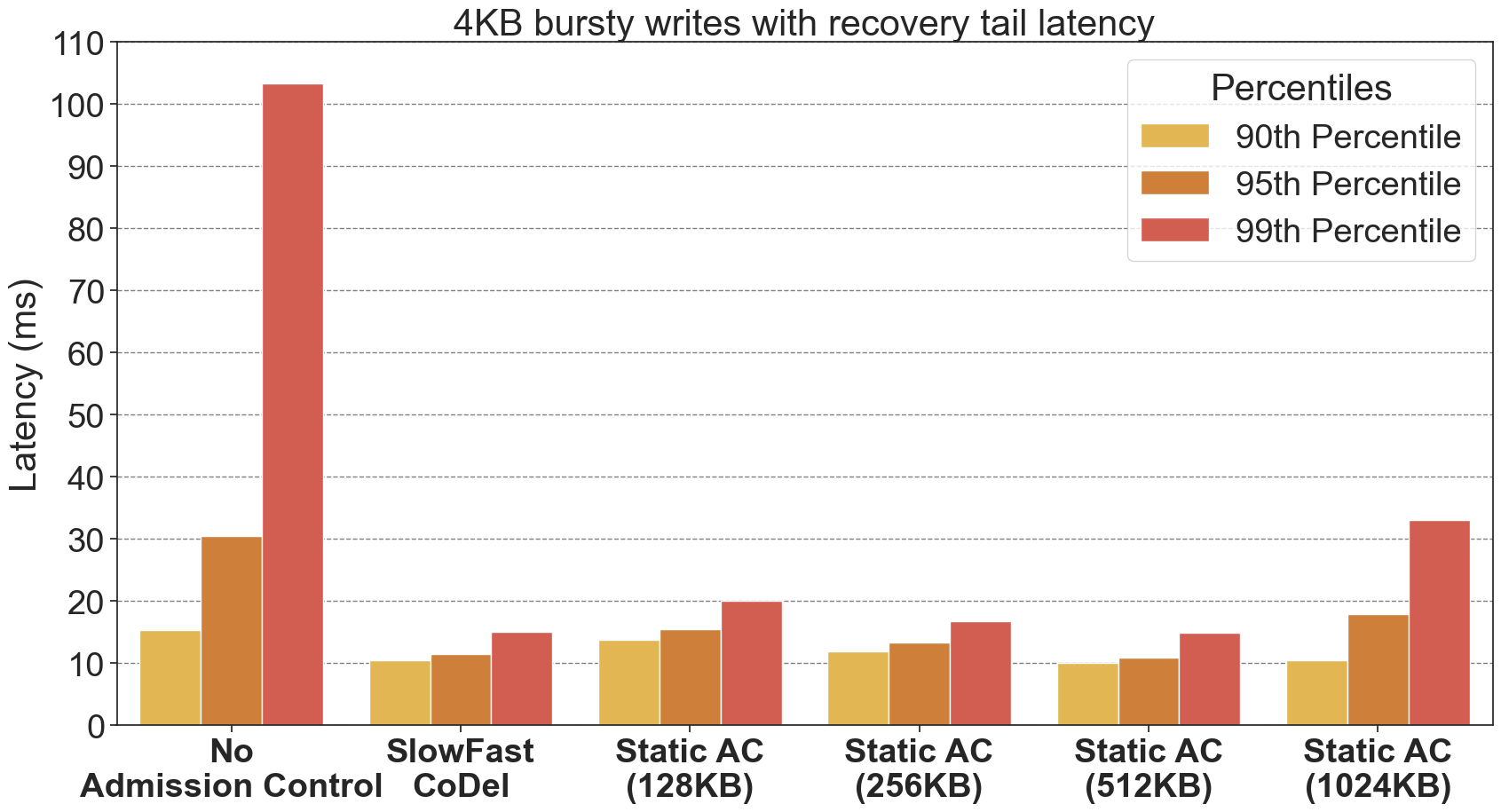}} 
  \vfill
  \subfloat[\textbf{32KB bursty writes} workload with recovery\label{fig:ac_bufferbloat_32k_64}]{\includegraphics[width=0.5\textwidth]{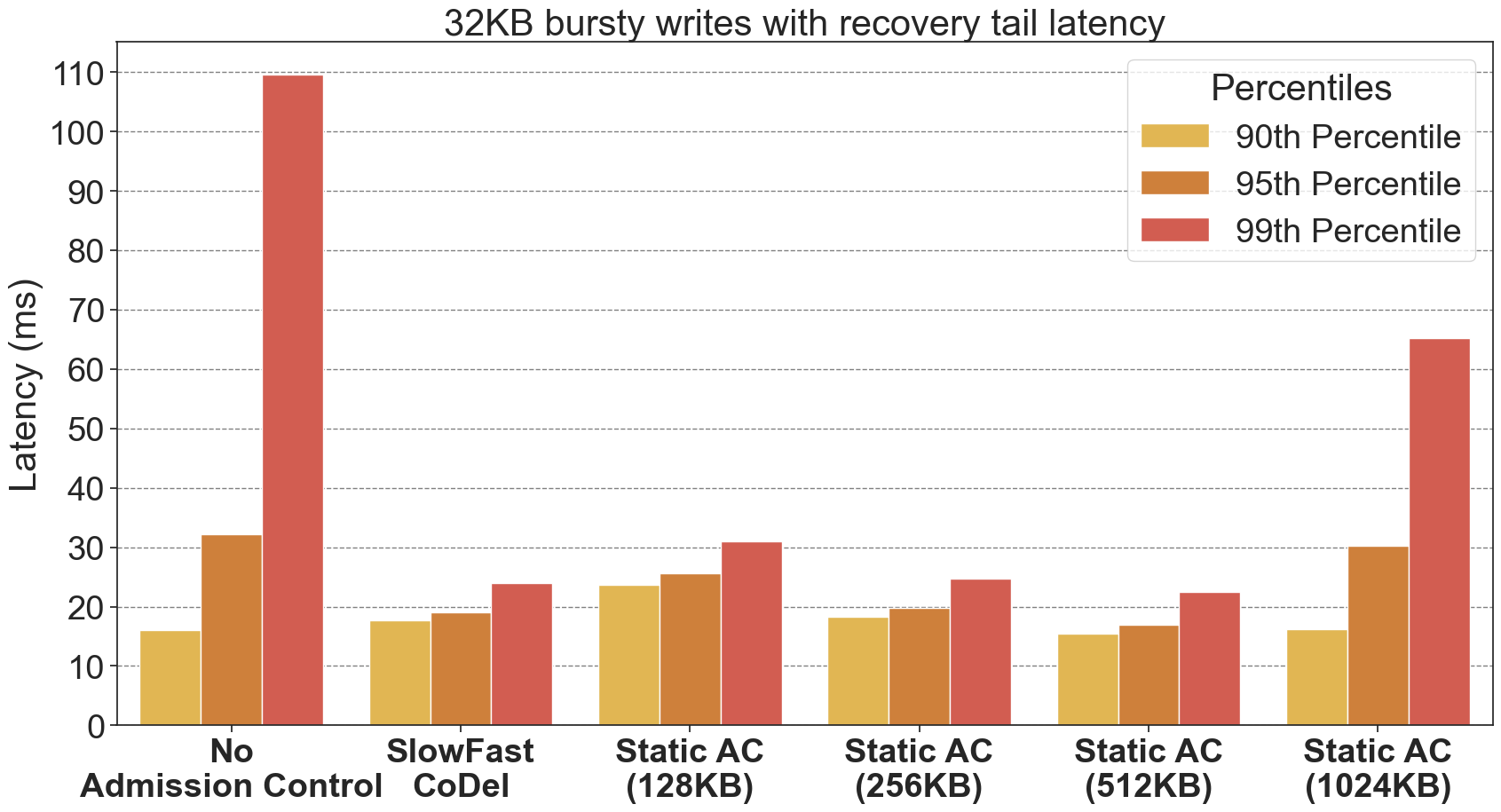}}
  \caption{Utilizing an admission control mitigates the bufferbloat impact on client.}
  \label{fig:recovery}
\end{figure}

On the other hand, Figure \ref{fig:recovery} shows the results of the same experiment when we utilize static admission control and SlowFast CoDel on the BlueStore in comparison to BlueStore without any admission control. Although recovery still increases the client tail latency when admission control is in place, the impact is limited and much lower than the setup without any admission control.

These results demonstrate the impact of bufferbloat on performance isolation in the Ceph storage system when no admission control is implemented on BlueStore. As we explained in previous sections, it is crucial to have proper admission control on the storage backend to mitigate and alleviate the bufferbloat.

Figure \ref{fig:recovery} shows the tail latency of 4KB and 32KB bursty writes with recovery. From left to right, the first set of bars shows the client workload's tail latency when the system has no admission control on the backend. The second set of bars is the results of the same workloads with our proposed SlowFast CoDel as an admission control in the BlueStore. The rest demonstrates the results for the BlueStore static throttle with budgets of 128KB, 256KB, 512KB, and 1024KB. As the figure depicts, using admission control can alleviate the bufferbloat impact.

In static admission control, when the budget is too low, like 128KB, tail latencies are slightly higher than 256KB and 512KB budget. The reason is that when we limit the BlueStore throttle budget to 128KB, the backend efficiency drops since the backend is starving. On the other hand, having a large budget, like 1024KB, leads to a higher bufferbloat impact on client latencies. The figure shows that SlowFast CoDel and BlueStore static throttle with 256KB and 512KB budget are more capable of mitigating bufferbloat.

These results demonstrate the need for an admission control mechanism in the storage backends. BlueStore utilizes static throttling with a fixed budget, which can mitigate the bufferbloat with the proper budget. However, finding the appropriate budget for BlueStore throttling requires performing benchmarks with the desired workload and hardware before deploying the storage backends. This can be very difficult in a dynamic cloud environment containing heterogeneous storage nodes and dynamic client workloads. The following section shows that static admission control mechanisms are inadequate in dynamic cloud storage environments.

\subsection{Necessity of an Adaptive Admission Control}
\label{need_adaptive}
In this section, we show our adaptive admission control, SlowFast CoDel, can achieve better performance than static admission controls while managing the bufferbloat.

To do so, we run the 4KB, 32KB, and 64KB continuous writes at a queue depth of 64 for 300 secs without any recovery. These workloads are continuous and non-bursty. Our goal is to evaluate performance of SlowFast CoDel and BlueStore static throttling.

\begin{figure}[h]
  \centering
  \subfloat[\textbf{4KB writes} workload at a queue depth of 64\label{fig:reg_4k}]{\includegraphics[width=0.5\textwidth]{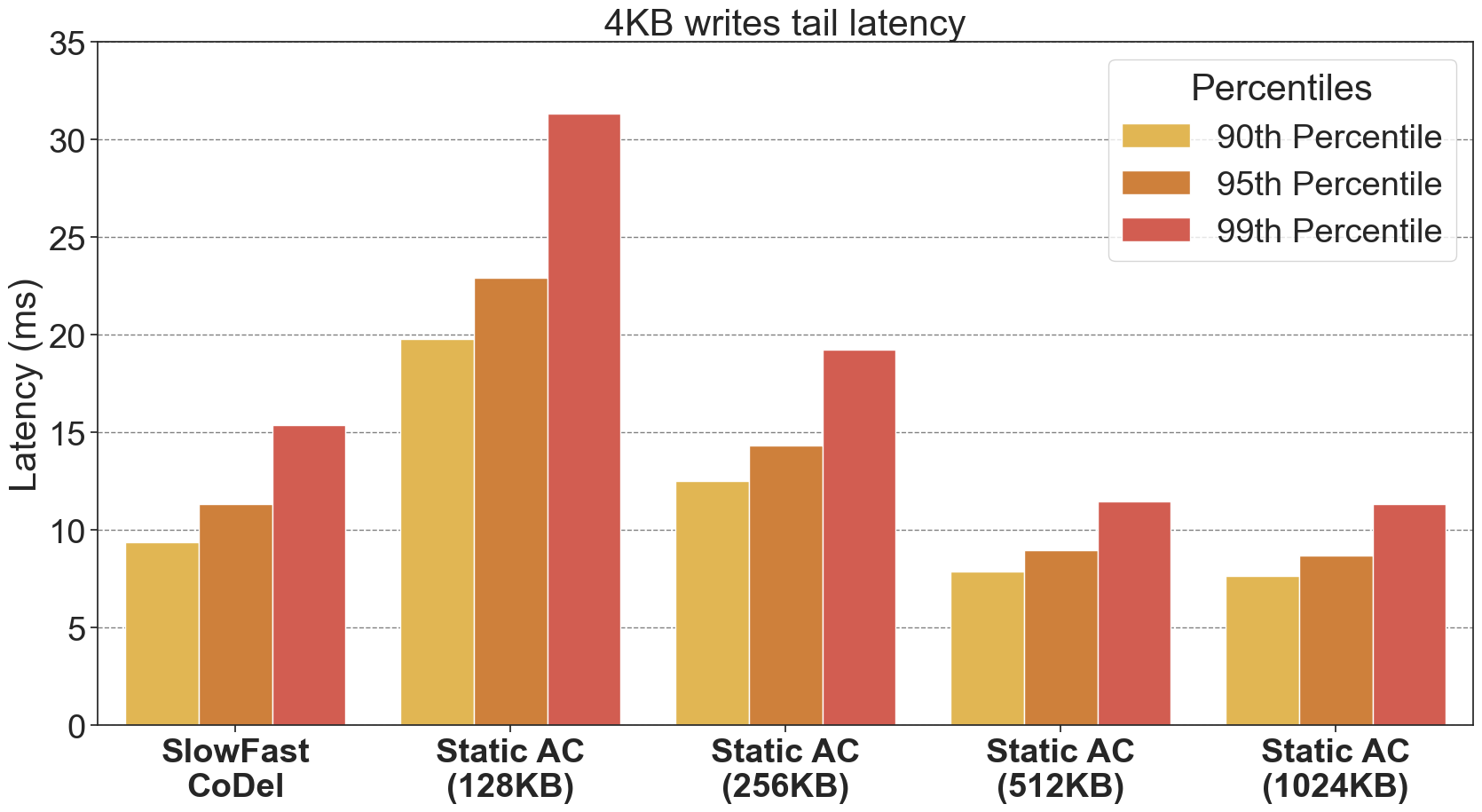}} 
  \vfill
  \subfloat[\textbf{32KB writes} workload at a queue depth of 64\label{fig:reg_32k}]{\includegraphics[width=0.5\textwidth]{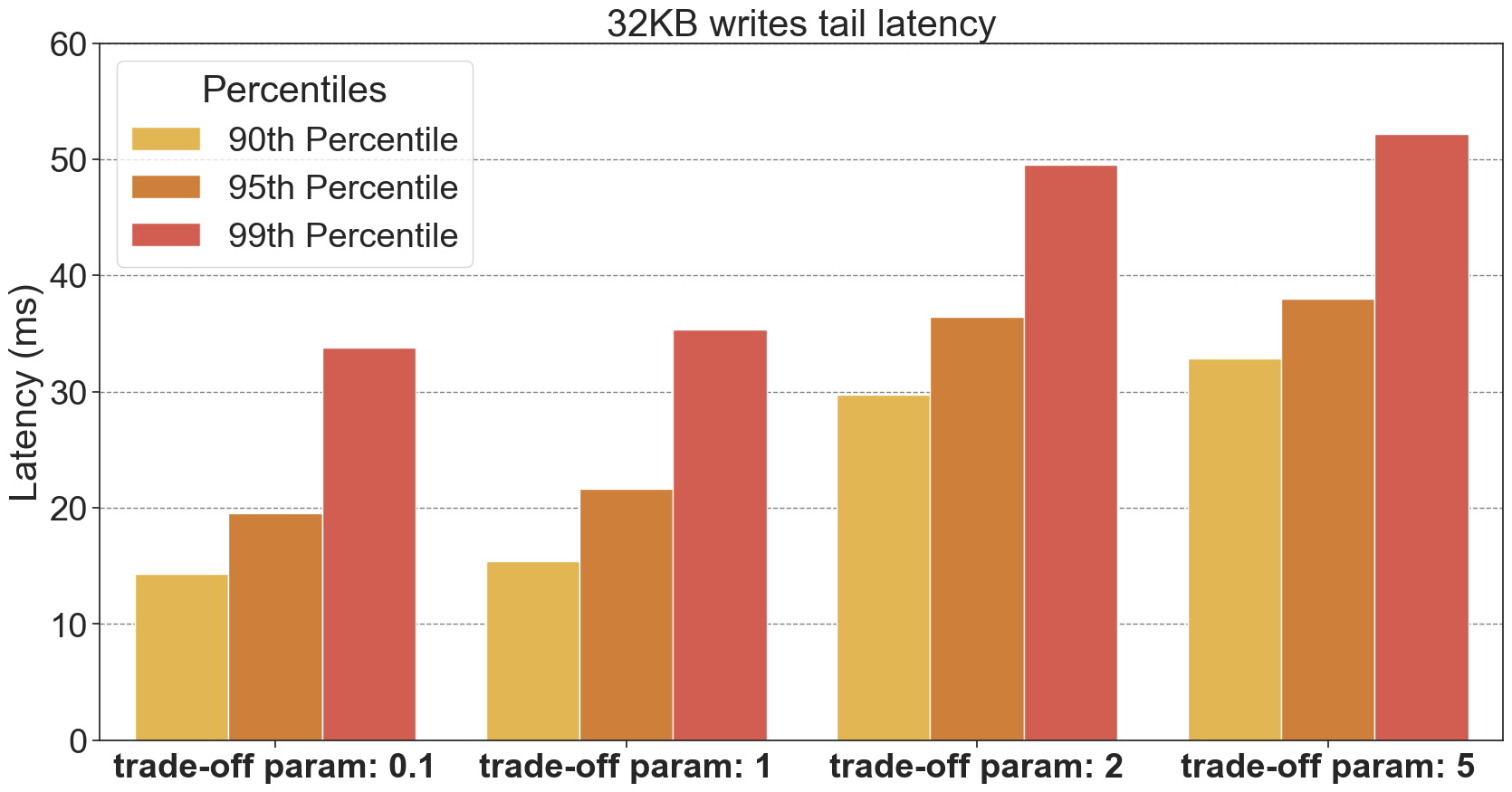}}
  \vfill
  \subfloat[\textbf{64KB writes} workload at a queue depth of 64\label{fig:reg_64k}]{\includegraphics[width=0.5\textwidth]{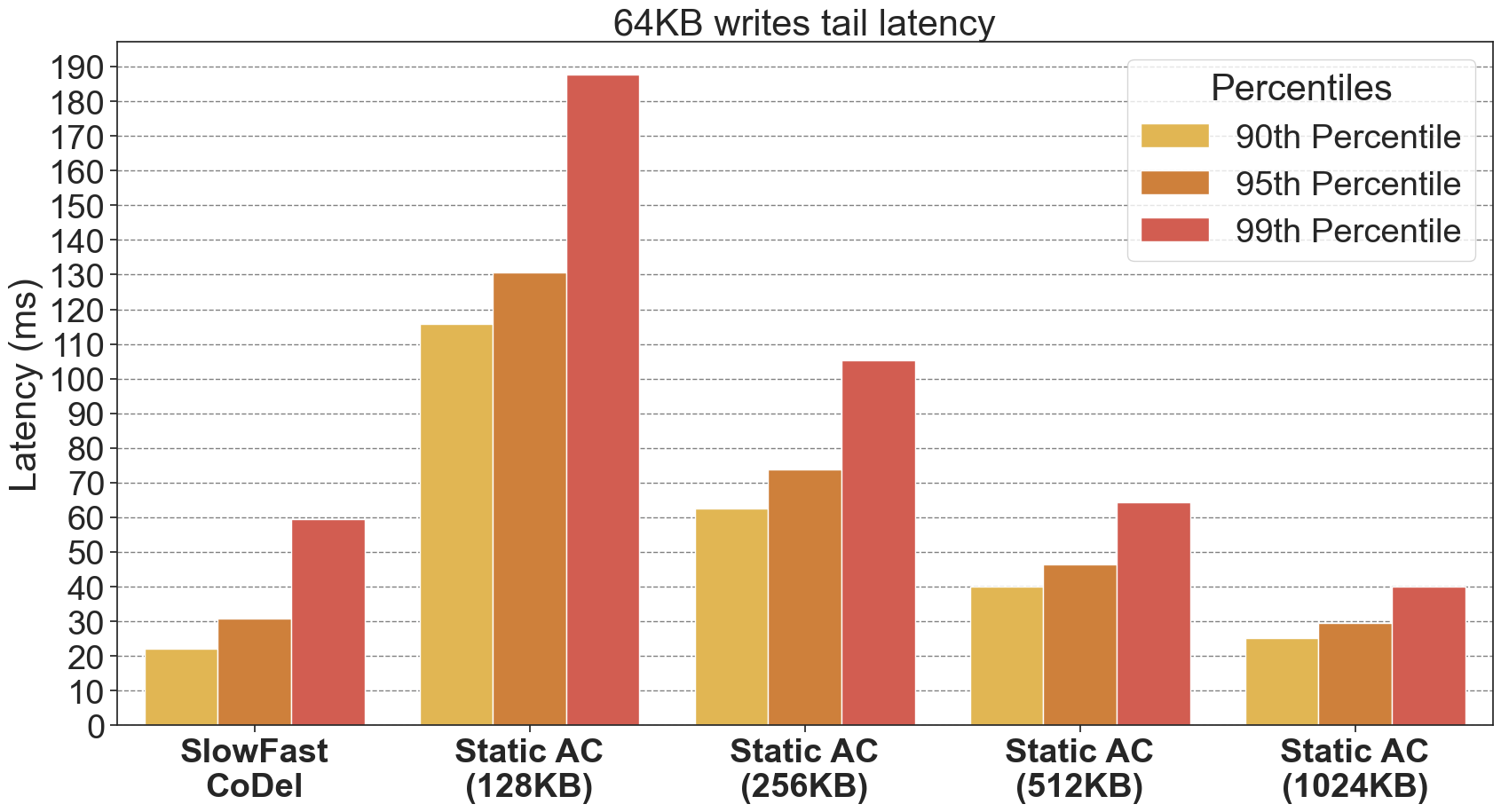}}
  \caption{Client tail latency comparison between static and adaptive admission control.}
  \label{fig:static_vs_adaptive}
\end{figure}

\begin{figure}[h]
  \centering
  \includegraphics[width=\linewidth]{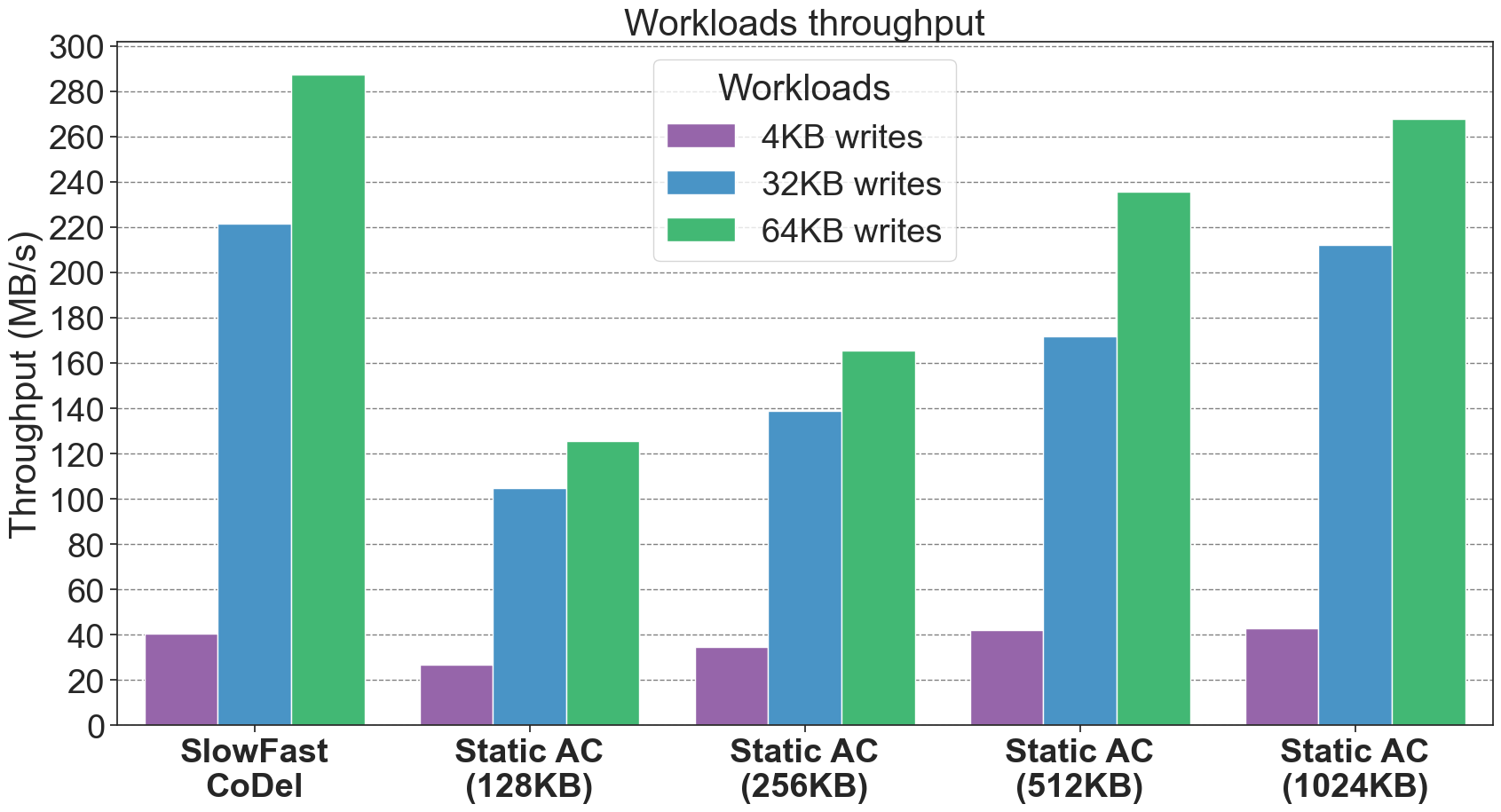}
  \caption{Client throughput comparison between static and adaptive admission control.}
  \label{fig:reg_thro}
\end{figure}

Figure \ref{fig:static_vs_adaptive} and Figure \ref{fig:reg_thro} show the results of these experiments. As Figure \ref{fig:static_vs_adaptive} demonstrates, a higher admission control budget, like 1024KB, can lead to better client tail latencies when there is no recovery workload. However, as we see in Figure \ref{fig:recovery} in the previous section, static admission control with 1024KB budget is not a good option to mitigate bufferbloat.

Figure \ref{fig:static_vs_adaptive} shows that SlowFast CoDel performs differently for different workload sizes due to its adaptive nature. SlowFast CoDel tries to minimize the latency while optimizing for the throughput. For 4KB workload, although SlowFast CoDel has higher tail latencies than static admission control with 512KB and 1024KB budget, it is more capable of mitigating the bufferbloat as it keeps the average admission control budget between 256KB and 512KB. For 32KB and 64KB workloads, SlowFast CoDel keeps the average budget between 512KB and 1024KB to achieve higher performance and mitigate bufferbloat better.

These results show that SlowFast CoDel adapts to different workloads and adjusts the admission budget according to the workload size. Moreover, Figure \ref{fig:reg_thro} shows throughputs of SlowFast CoDel and static admission controls for different workload sizes. The figure shows that SlowFast CoDel can outperform static admission controls for 32KB and 64KB workloads as it can dynamically adjust the admission budget to optimize throughput.

\subsection{Trade-off Parameter in Target Adjusting Loop}
\label{sens}
In this section, we repeat the experiments from section \ref{need_adaptive} for different values of the Trade-off parameter. Although the Trade-off parameter is basically the slope of the Throughput-Latency curve and has no unit, the units of the y-axis and x-axis of this curve are important. In our implementation, the y-axis is the backend throughput in MB/s, and the x-axis is the backend latency in ms. In this experiment, we pick values of 0.1, 1, 2, and 5 for the Trade-off parameter. 

\begin{figure}[h]
  \centering
  \subfloat[\textbf{4KB writes} workload at a queue depth of 64\label{fig:trade_off_4k}]{\includegraphics[width=0.5\textwidth]{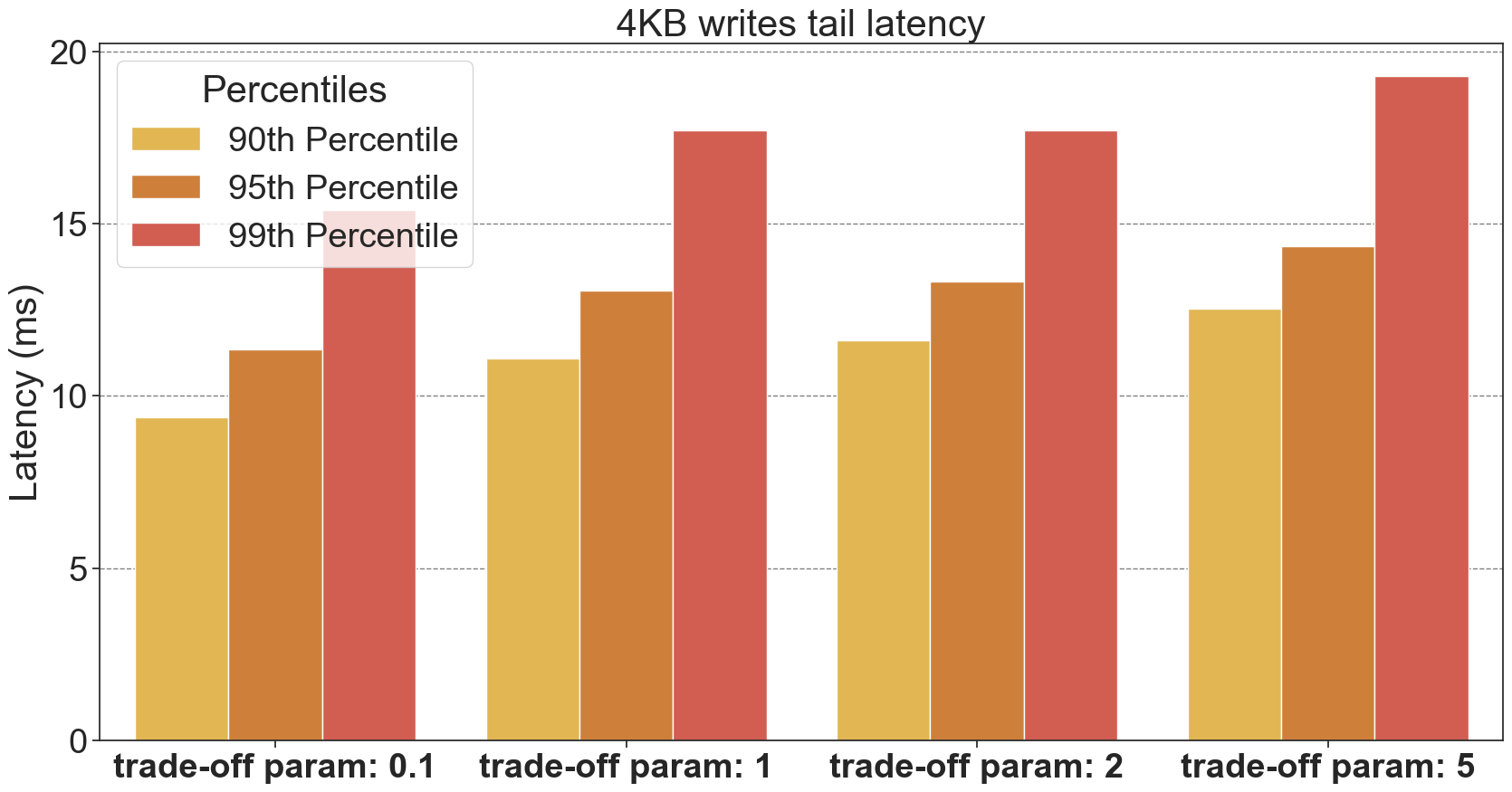}} 
  \vfill
  \subfloat[\textbf{32KB writes} workload at a queue depth of 64\label{fig:trade_off_32k}]{\includegraphics[width=0.5\textwidth]{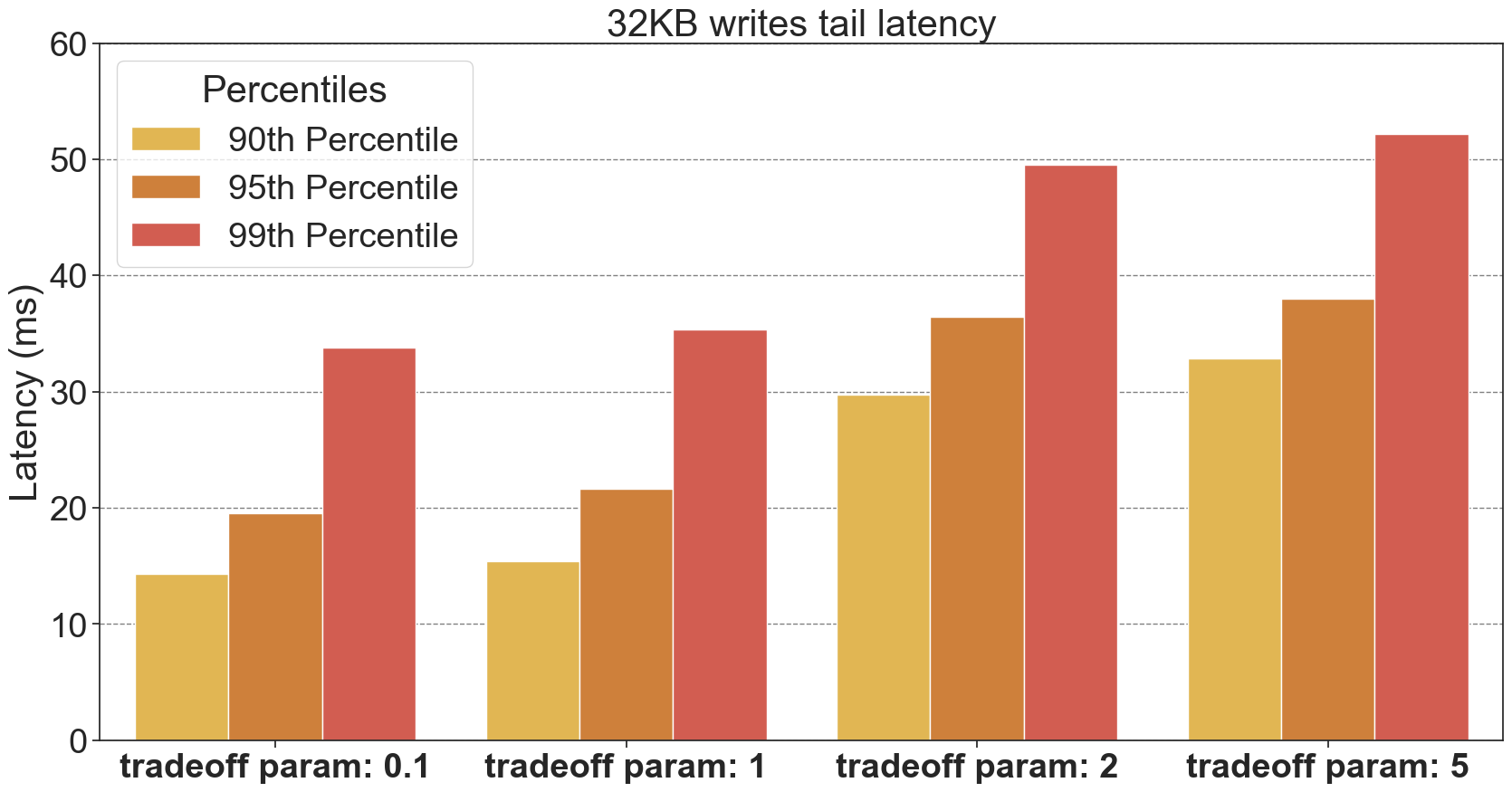}}
  \vfill
  \subfloat[\textbf{64KB writes} workload at a queue depth of 64\label{fig:trade_off_64k}]{\includegraphics[width=0.5\textwidth]{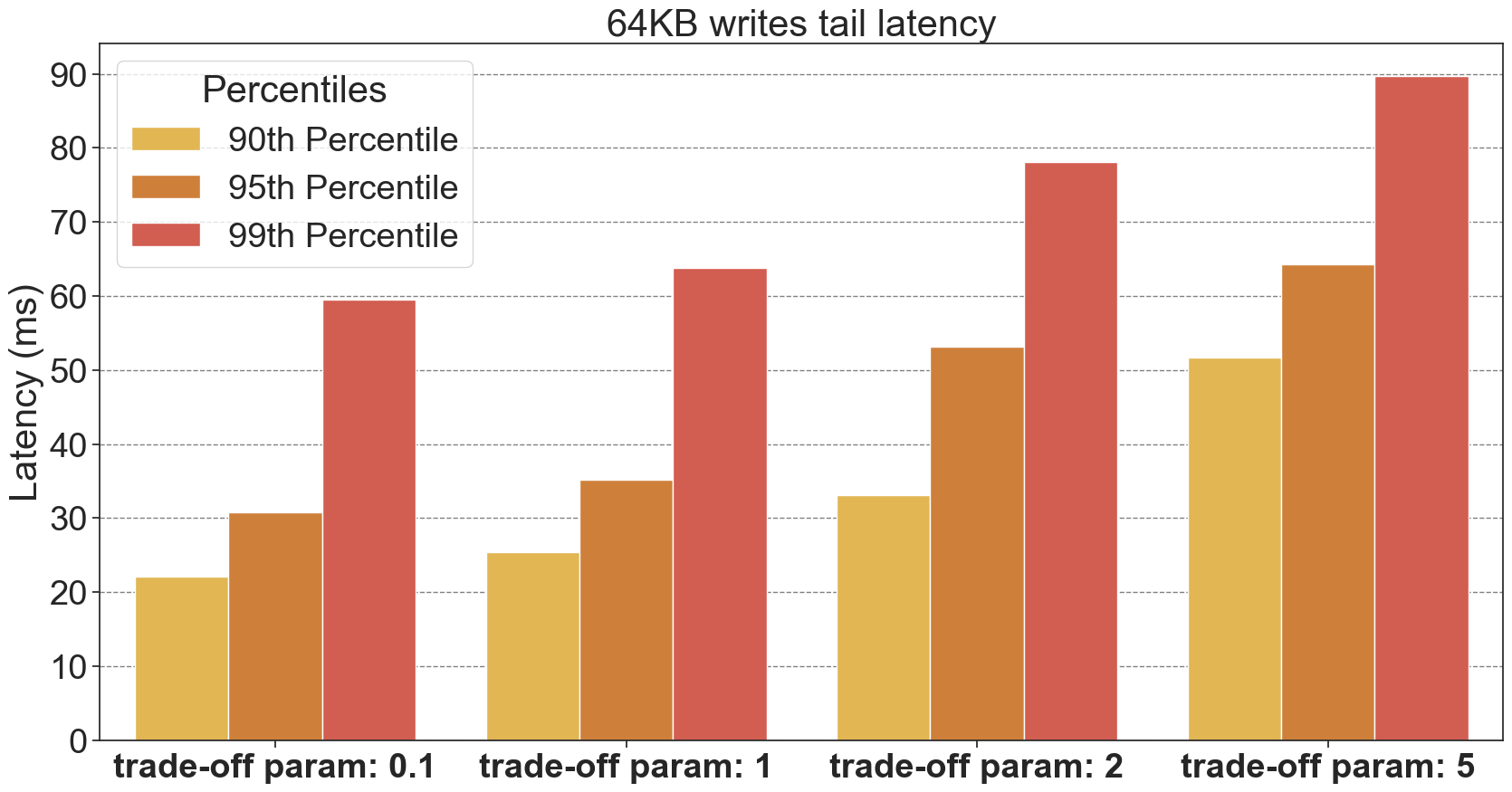}}
  \caption{Client tail latency comparison between SlowFast CoDel with different Trade-off parameter.}
  \label{fig:codel_sens}
\end{figure}

\begin{figure}[H]
  \centering
  \includegraphics[width=\linewidth]{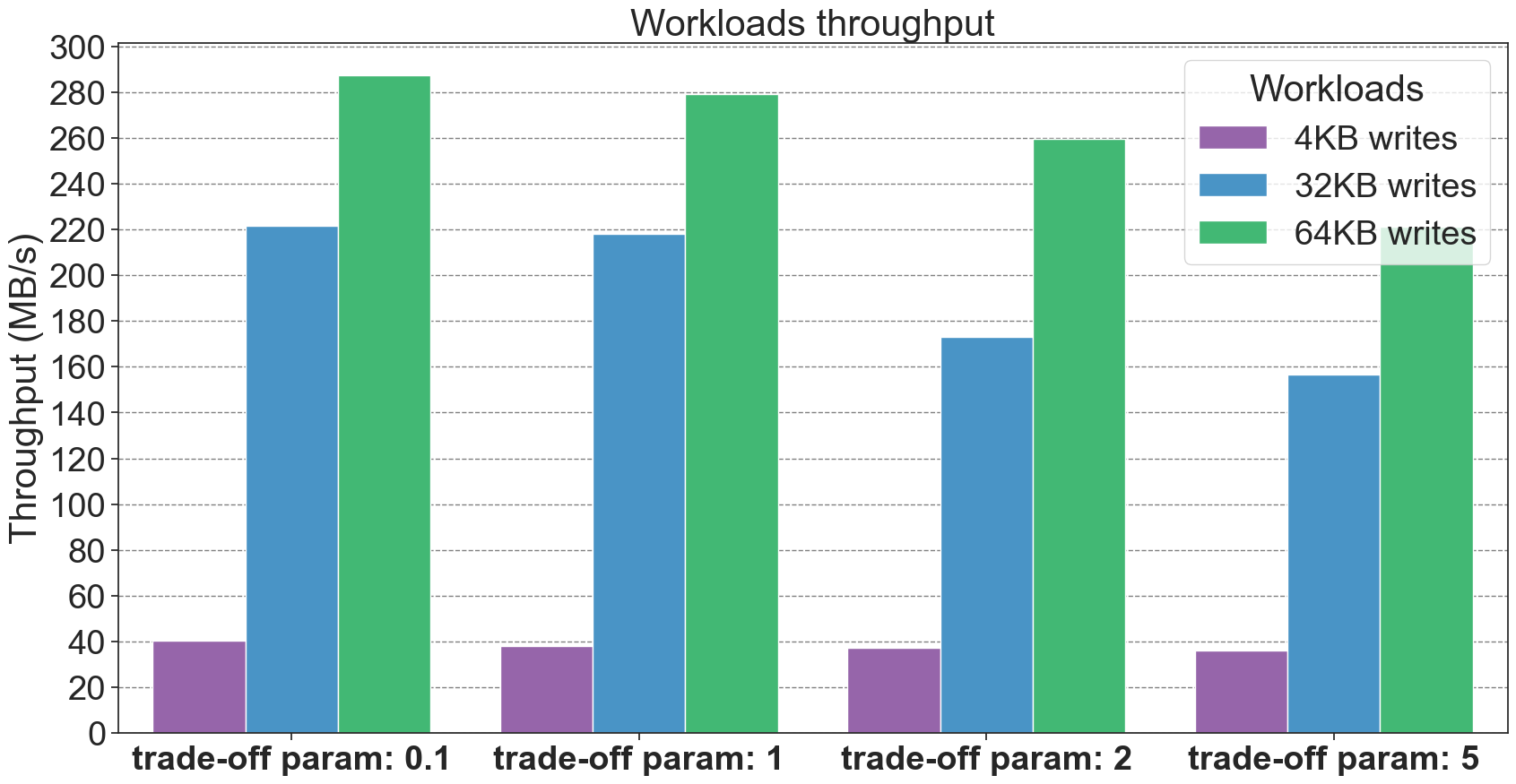}
  \caption{Client throughput comparison between SlowFast CoDel with different Trade-off parameter.}
  \label{fig:codel_sens_thro}
\end{figure}

Figure \ref{fig:codel_sens} shows the client tail latencies and Figure \ref{fig:codel_sens_thro} shows the client throughput for SlowFast CoDel with different Trade-off parameters. As Figures \ref{fig:codel_sens} and \ref{fig:codel_sens_thro} demonstrate, selecting a high Trade-off parameter leads to latency increase and throughput loss. The reason is that a higher Trade-off parameter limits the backend admission control and makes the backend starve. Moreover, the results show that SlowFast CoDel with any of the tested Trade-off parameters performs consistently on different workloads.

\section{Conclusion}
\label{conclusion}
To the best of our knowledge, we identify the bufferbloat as a schedulability issue for the first time and investigate the impact of the bufferbloat on performance isolation scheduling in cloud storage QoS. The results of our experiments in section \ref{schedulability_impact} show that the bufferbloat leads to priority inversion and impact the client tail latencies.

Then, we focus on finding a solution that can be utilized easily without any change in system architecture and high-level QoS and scheduling mechanisms. In order to do that, we show that utilizing an admission control mechanism in the storage backend can mitigate the bufferbloat problems. Moreover, we demonstrate that static admission controls with fixed admission budgets are inadequate in the face of dynamic workloads of cloud environments.

To make our case, we propose an adaptive admission control called SlowFast CoDel which utilizes a dual-loop control mechanism. The inner loop of SlowFast CoDel is a modified CoDel algorithm that adjusts the queueing budget of the backend based on the backend latency. Then, an outer loop optimizes and trade-off the latency and throughput by adjusting the inner loop parameter target latency. Section \ref{need_adaptive} shows that SlowFast CoDel can mitigate bufferbloat for multiple workload sizes.

The main contributions of this paper are identifying and demonstrating the impact of bufferbloat on the schedulability of cloud storage systems and acknowledging the necessity of admission control in the storage backend to mitigate this impact. This paper focuses on identifying the impact of bufferbloat on schedulability and the necessity of admission control to mitigate bufferbloat. We propose SlowFast CoDel as a starting point for researching adaptive admission control in the storage backends. While the SlowFast CoDel is an important contribution, discussing and analyzing it in detail would not fit in the scope of this paper and detract from the paper's primary focus. In future works, we plan to analyze and discuss SlowFast CoDel in detail. 
\section{Acknowledgements} 
\label{acknowledge}
This work has been made possible by the Center for Research in Open Source Software at UC Santa Cruz (cross.ucsc.edu) and supported by the National Science Foundation under Cooperative Agreement OAC-1836650.

%
%
\bibliographystyle{splncs04}
\bibliography{refrences}

\end{document}